# MONITORING SURFACE DEFORMATION OVER OILFIELD USING MT-INSAR AND PRODUCTION WELL DATA

*Sarah Narges Fatholahi[1], Hongjie He[1], Lanying Wang[1], Awase Syed[1], Jonathan Li[1,2*]*

[1]Department of Geography and Environmental Management, University of Waterloo, Waterloo, ON N2L 3G1, Canada
[2]Department of Systems Design Engineering, University of Waterloo, Waterloo, ON N2L 3G1, Canada
nfatholahi@uwaterloo.ca (S.N. Fatholahi), h69he@uwaterloo.ca (H. He), lanying.wang@uwaterloo.ca (L. Wang) and awase008@gmail.com (A. Syed)
*Corresponding authors: junli@uwaterloo.ca (J. Li)

**ABSTRACT**

Surface displacements associated with the average subsidence due to hydrocarbon exploitation in southwest of Iran which has a long history in oil production, can lead to significant damages to surface and subsurface structures, and requires serious consideration. In this study, the Small BAseline Subset (SBAS) approach, which is a multi-temporal Interferometric Synthetic Aperture Radar (InSAR) algorithm was employed to resolve ground deformation in the Marun region, Iran. A total of 22 interferograms were generated using 10 Envisat ASAR images. The mean velocity map obtained in the Line-Of-Sight (LOS) direction of satellite to the ground reveals the maximum subsidence on order of 13.5 mm per year over the field due to both tectonic and non-tectonic features. In order to assess the effect of non-tectonic features such as petroleum extraction on ground surface displacement, the results of InSAR have been compared with the oil production rate, which have shown a good agreement.

## 1. INTRODUCTION

Oilfields are frequently exploited for oil production, which leads to instability in the ground surface followed by land subsidence. Fluid exploitation and injection activities which cause surface deformation over oil fields can lead to geohazards [1]. Therefore, the assessment of ground surface displacement due to oil exploitation activities is crucial for efficient production of oilfields, and safety assessments of surrounding infrastructure [2].

Land loss is particularly noticeable on coastal areas where a slight decline of the land surface may lead to permanent inundation; and first occurred in Goose Creek oil field along the Texas Gulf coast, USA in which roads and coastlines were totally damaged [3]. Wilmington field in Long Beach, California was another dramatic case of land subsidence due to oil and gas extraction which experienced almost 9 m of subsidence and caused structural damages to surface facilities (e. g. railroads and bridges) [4].

Techniques based on revisiting ground-based measurements, such as in-situ geodetic data, precise leveling and Global Navigation Satellite System (GNSS) generally offer precise data well established by the authorities in charge of risk management. However, these methods which require large number of individual observations as well as ground access, not only laborious and high cost, but also restricted to measure variations in the locations of limited set of benchmarks.

Interferometric Synthetic Aperture Radar (InSAR) is an effective remote sensing technique, which has confirmed the accurate assessment of ground surface movements at mm level using the phase information of SAR images [5]. This method as a globally accepted practice for long-term measurement of ground surface movements enables the mapping of large areas at low cost. InSAR technique has been successfully employed for detecting ground surface displacement over oil and geothermal fields such as the Helliishdi geothermal field [6], the West Texas oilfield [7], and the Cushing oilfield in East Central Oklahoma [8].

The goal of this study is to quantitatively assess the ground surface displacement rate (subsidence and uplift) over a giant oilfield in Iran to investigate the corresponding subsidence source. For this purpose, Small Baseline Subset approach was applied to detect deformation velocities in Marun oilfield.

## 2. MATERIALS AND METHODS

### 2.1 Study area

The Marun anticline is one of the supergiant oilfields located in southwest region of Zagros fold-thrust belt in

Iran, which has been formed as a result of the continental collision between the Arabian plate and Iranian block (Fig. 1) [9]. Its anticline is obvious in the west and central parts, in the NW-SE direction. This field has an asymmetric structure with an average slope of 45°- 60° in southwest flank and a 25° - 45° in the northeast flank. It has also a twist in the middle of the structure, that the direction of field changes nearly 20° from the N45°W direction in northwest to N65°W of Southwest. In Marun oilfield, the slope variations in the flanks are different due to two significant tectonic events including the folding which is the main effect of Zagros orogeny and flexion generated by compressive force. As a result, the radius of curvature of various sections in the structure is different.

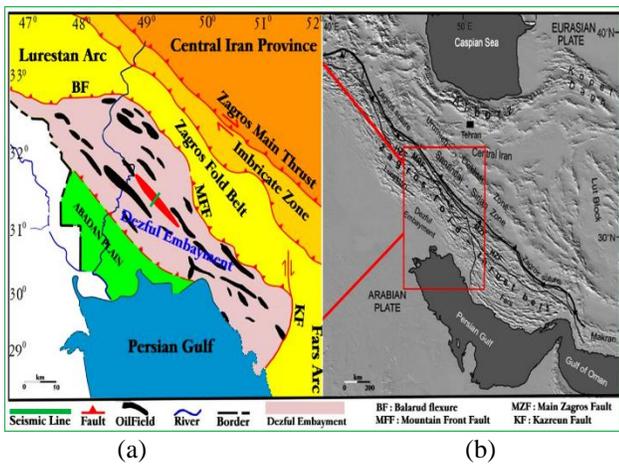

(a) (b)

Fig. 1. (a) Location map and main structural features related to Zagros fold-thrust belt. (b) Locality map of Marun oilfield (red) (modified from [10])

## 2.2 Methods

Temporal and geometrical baseline decorrelations as well as imprecision of the reference digital elevation mode (DEM) are the key shortcomings of the measurement of slow movements in standard differential interferometry SAR (D-InSAR) method in subsidence-prone regions. Recent developments in multi-temporal InSAR (MT-InSAR) algorithms have raised the capability to detect slow deformations with millimeter precision by obtaining the early signal of hazard that is helpful in forecasting potential subsidence. The Small Baseline Subset (SBAS) technique identifies coherent pixels with phase stability over a specific observation period. This approach is based on multiple-master interferograms and operates with limited spatial baseline interferograms and short time intervals to resolve de-correlations by increasing spatial and temporal sampling and coherent areas. In this research, 10 ASAR images acquired by the ENVISAT satellite from European Space Agency (ESA) were used in a descending geometry mode, track 149 from 2003 to 2006 (Table 1).

Table 1. Acquisition date of Envisat C-band SAR images

| Date of acquisition (yyyymmdd) | Orbit number |
|---|---|
| 20030926 | 8219 |
| 20031205 | 9221 |
| 20040109 | 9722 |
| 20040213 | 10223 |
| 20040423 | 11225 |
| 20040528 | 11726 |
| 20040806 | 12728 |
| 20050513 | 16736 |
| 20050722 | 17738 |
| 20050826 | 18239 |

The conversion of raw to the Single Look Complex (SLC) SAR images was carried out using orbital and sensor calibration data in the Repeat Orbit Interferometry PACkage (ROIPAC) – an open access software. Then, the co-registration analysis was done using DORIS software [11] based on the image acquired on April 23th, 2004. SBAS processing was also implemented by applying the Stanford Method for Persistent Scatterers (StaMPS) software [12][13] to build multiple masters interferograms. Fig. 2 shows the workflow of the SBAS method.

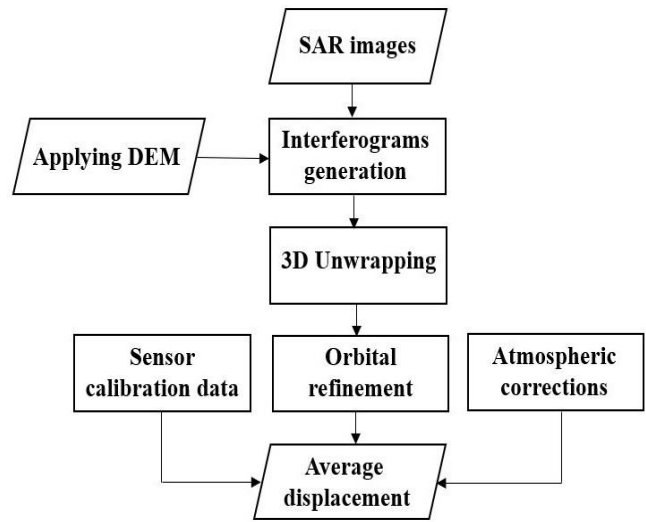

Fig. 2. Workflow of the SBAS approach

## 3. RESULTS AND EVALUATION

We created the network with 22 differential interferograms (Table 2) with minimal spatial, temporal, and Doppler baselines. The interferograms were generated using the repeat-pass technique implemented in Delft object-oriented radar interferometric software (DORIS).

The Shuttle Radar Topography Mission digital elevation model (SRTM DEM) with 3-arcsecond geographical resolution was used to remove the topographic phase. Fig. 3 represents the SBAS network derived from the time series analysis. Fig. 4 shows the mean velocity map obtained through InSAR time series analysis, which is in the Line-Of-Sight (LOS) direction of satellite to the ground. As shown in Fig. 4, the subsidence with the maximum rate of approximately 13.5 mm/yr was happening over the Marun region during 2003 to 2006.

Table 1. Characteristics of processed interferograms

| Master date (yyyymmdd) | Slave date (yyyymmdd) | Bperp (m) |
| --- | --- | --- |
| 20030926 | 20031205 | 35 |
| 20030926 | 20040213 | 34.8 |
| 20030926 | 20040423 | 378 |
| 20030926 | 20040806 | 175.7 |
| 20031205 | 20040213 | 343.2 |
| 20031205 | 20040423 | 343 |
| 20031205 | 20040806 | 140.7 |
| 20040109 | 20040423 | 242.2 |
| 20040109 | 20041015 | 121.2 |
| 20040109 | 20050513 | 93.7 |
| 20040109 | 20050722 | 153.9 |
| 20040109 | 20050826 | 48.1 |
| 20040213 | 20040423 | 343.2 |
| 20040213 | 20040806 | 140.9 |
| 20040423 | 20040806 | -202.3 |
| 20040423 | 20041015 | 393.4 |
| 20041015 | 20050513 | -27.5 |
| 20041015 | 20050722 | 32.7 |
| 20041015 | 20050826 | -73.1 |
| 20050513 | 20050722 | 60.2 |
| 20050513 | 20050826 | -45.6 |
| 20050722 | 20050826 | -105.8 |

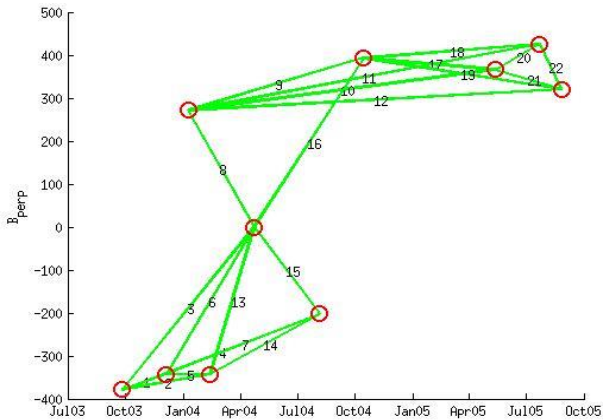

Fig. 3. The SBAS network derived from the time series analysis: vertical and horizontal axes indicate the perpendicular and temporal baselines, respectively.

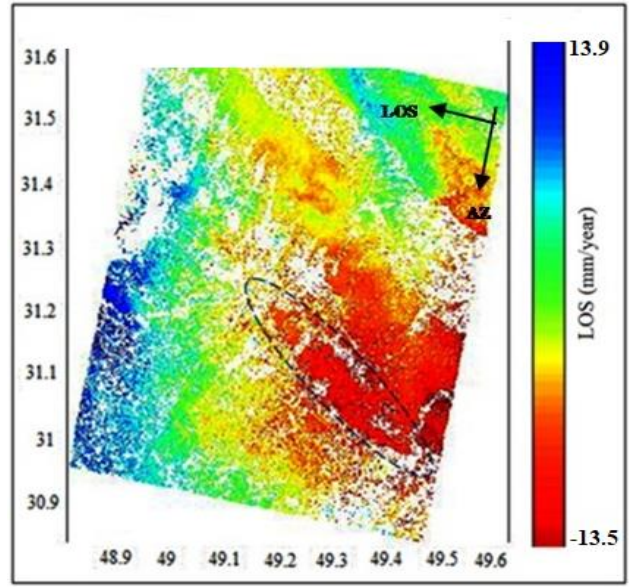

Fig. 4. Mean velocity map of InSAR time series analysis during 2003 and 2006

Sampled oil-well measurements from two producing wells number 78 and 166 in the central part of the field were considered and their monthly production rates were specified at the same time as interferometric pairs. This dataset allows us to assess the potential of non-tectonic impacts such as petroleum extraction on surface displacements and in particular, the relationship between both deformation and oil production rates.

The production patterns and surface behaviours in both wells 78 and 166 are shown in Fig. 5. These figures demonstrate some striking similarities between these two wells from 2003 to 2006. Both wells have experienced considerable fluctuations in exploitation within the same duration, with some lows during September 2003 to August 2004, followed by December 2003 to February 2004. And some rises up to January 2004 and May 2005. For well 78, with increased production during September 2003 to August 2004, the surface elevation decreased to about -5 mm. The production rate in this well rose from about $5 \times 10^5$ barrel between December 2003 and February 2004 to just $3 \times 10^6$ barrel between January 2004 and May 2005. Then, the rate of production plunged to a lowest level between February and April 2004. The production rates gradually increased by $2 \times 10^6$ barrels in the following time periods, and then declined to the minimum amount of production in the next interferogram. Also, the well 166 had similar production rate as well 78.

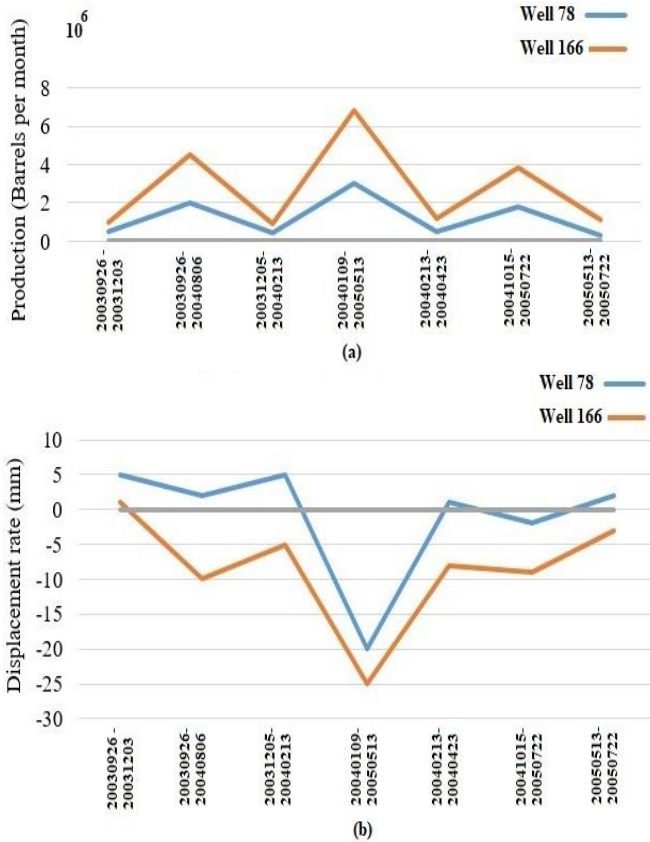

Fig. 5. (a) Monthly production rate and (b) Time-series plots of displacement

## 4. CONCLUSION

In this study, time series analysis of SAR data has been applied to resolve ground deformation in the Marun oilfield located in the southwest of Iran. The results reveal the maximum subsidence in the area was to an order of 13.5 mm per year over the span of three years. The monthly production rate of sampled wells is then combined with surface deformation rate. The displacement activity in the central part of the field is consistent with the pace of oil production.

The variation obtained from ground movements are too complex to be expressed for one or two reasons. However, results show that deformation caused by subsidence in Marun region is influenced by the combined effects of tectonic activity, subsurface pressure reduction, hydraulic fractures, physical properties of reservoir's rocks, and oil production rates.

## 5. REFERENCES

[1] Hu, B., Li, H., Zhang, X., and Fang, L., 2020. Oil and Gas Mining Deformation Monitoring and Assessments of Disaster: Using Interferometric Synthetic Aperture Radar Technology. *IEEE Geoscience Remote Sensing Magazine*, 8 (2), 108-134.

[2] Togaibekov, A.Z., 2020. Monitoring of oil-production-induced subsidence and uplif. MSc Thesis in Geophysics, Massachusetts Institute of Technology, https://hdl.handle.net/1721.1/127149.

[3] Nagel, N., 2001. Compaction and subsidence issues within the petroleum industry: From Wilmington to Ekofisk and beyond. *Physics and Chemistry of the Earth* Pt. A, 26, 3–14.

[4] Poland, J.F., and Davis, G.H., 1969. Land subsidence due to withdrawal of fluids. *Reviews in Engineering Geology* 2, 187-269.

[5] Li, D., Hou, X., Song, Y., Zhang, Y., and Wang, C., 2020. Ground subsidence analysis in Tianjin (China) based on Sentinel-1A data using MT-InSAR methods. *Applied Science* 10, 5514.

[6] Juncu, D., Árnadóttir, T., Geirsson, H., Guðmundsson, G.B., Lund, B., Gunnarsson, G., Michalczewska, K., 2020. Injection-induced surface deformation and seismicity at the Hellisheidi geothermal field, Iceland. *Journal of Volcanology and Geothermal Research* 391, 106337.

[7] Yang, Q., Zhao, W., Dixon, T.H., Amelung, F., Han, W.S., Li, P., 2015. InSAR monitoring of ground deformation due to $CO_2$ injection at an enhanced oil recovery site, West Texas. *International Journal of Greenhouse Gas Control*. 41, 20–28.

[8] Loesch, E., Sagan, V., 2018. SBAS analysis of induced ground surface deformation from wastewater injection in east central Oklahoma, USA. *Remote Sensing* 10, 283–299.

[9] Sherkati, S., and J. Letouzey. 2004. Variation of structural style and basin evolution in the central Zagros (Izeh zone and Dezful Embayment), Iran. *Marine and Petroleum Geology* 21, 535e554.

[10] Fakhari, M., Axen, G.J., Horton, B.K., Hassanzadeh, J., Amini, A., 2008. Revised age of proximal deposits in the Zagros foreland basin and implications for Cenozoic evaluation of the High Zagros. *Tectonophy*. 451, 170-185.

[11] Kampes, B., and Usai, S., 1999. Doris: The Delft object-oriented Radar Interferometric software. *In: Proceedings ITC 2nd ORS Symposium.*

[12] Hooper, A., 2008. A multi-temporal InSAR method incorporating both persistent scatterer and small baseline approaches. *Geophysical Research Letters* 2008, 35.

[13] Hooper, A., Segall, P., and Zebker, H., 2007. Persistent scatterer interferometric synthetic aperture radar for crustal deformation analysis. *Journal of Geophysical Research* 2007, 112.